\documentclass[aps,prl,reprint,superscriptaddress]{revtex4-1}

\usepackage{graphicx}
\usepackage{braket}
\usepackage{amsmath,amssymb,bm}
\usepackage{xcolor}
\usepackage{wasysym}

\begin{document}
\title{Decay of an isolated monopole into a Dirac monopole configuration}

\author{K. Tiurev}
\author{E. Ruokokoski}
\author{H. M\"akel\"a}
\affiliation{QCD Labs, COMP Centre of Excellence, Department of Applied Physics, Aalto University,
P.O. Box 13500, FI-00076 Aalto, Finland}
\author{D. S. Hall}
\affiliation{Department of Physics and Astronomy, Amherst College, Amherst, MA 01002-5000, USA}
\author{M. M\"ott\"onen}
\affiliation{QCD Labs, COMP Centre of Excellence, Department of Applied Physics, Aalto University,
P.O. Box 13500, FI-00076 Aalto, Finland}
\date{\today}

\begin{abstract}
We study numerically the detailed structure and decay dynamics of isolated monopoles in conditions similar to those of their recent experimental discovery. We find that the core of a monopole in the polar phase of a spin-1 Bose--Einstein condensate contains a small half-quantum vortex ring. Well after the creation of the monopole, we observe a dynamical quantum phase transition that destroys the polar phase. Strikingly, the resulting ferromagnetic order parameter exhibits a Dirac monopole in its synthetic magnetic field.
\end{abstract}

\maketitle

\emph{Introduction}---The significant roles played by topological defects in nature and their appearance in 
various physical contexts~\cite{Mermin1979,Nakahara2003} have sparked numerous theoretical and experimental studies.
The precise control over experimental parameters and the ability to image the quantum mechanical order parameter directly render Bose--Einstein condensates (BECs) a unique platform to create and observe different types of topological defects. In particular, BECs with spin degrees of freedom may host a rich variety of defects due to many possible order parameter manifolds and symmetries ~\cite{Stoof2001,Mizushima2002,Ruostekoski2003,Kawaguchi2008,Kobayashi2009,Pietila2009,Kawakami2011,Borgh2012}. In these systems, topological defects can either be created in a deterministic manner using precisely controlled magnetic and laser fields~\cite{Matthews1999,Leanhardt2002,Ogawa2002,Choi2012a,Ray2014}, or they can form spontaneously, for example when the condensate is rapidly quenched a through quantum phase transition~\cite{Weiler2008, Sadler2006}. The experimentally realized topological structures in BECs to date include singly and multiply quantized vortices~\cite{Matthews1999, Madison2000, Leanhardt2002,Isoshima2007,Fetter2009}, solitons and vortex rings~\cite{Anderson2001}, skyrmions~\cite{Choi2012,Choi2012a}, polar core vortices~\cite{Sadler2006}, coreless vortices~\cite{Leanhardt2003, Leslie2009}, vortex-antivortex superpositions~\cite{Wright2009}, solitonic vortices~\cite{Donadello2014} and monopoles~\cite{Ray2014, Ray2015}.


The Dirac monopole configuration created in Ref.~\cite{Ray2014} is an analogue of the classical magnetic point charge considered by Dirac in the context of quantum mechanics~\cite{Dirac1931}. It manifests itself as a point-like singularity in the so-called synthetic magnetic field~\cite{Kawaguchi2012},
which is an effective gauge field for the scalar part of the order parameter arising naturally from its spin degrees of freedom. In agreement with Dirac's original work~\cite{Dirac1931}, this kind of monopole induces in the condensate order parameter a nodal vortex line with vanishing particle density extending from the location of the monopole to the boundary of the atom cloud. Thus the ferromagnetic order parameter supporting the Dirac monopole is energetically and dynamically reminiscent of a line defect. Critically, there is no topological point defect in the order parameter itself, as it is in a configuration topologically equivalent to the ground state. Indeed, the second homotopy group~\cite{Nakahara2003} for the ferromagnetic order parameter space is trivial and topological point defects are not permitted. Point defects may exist in the polar phase of a spin-1 condensate, however, as the second homotopy group for the polar order parameter space, $G_{p}=[S^2\times U(1)]/\mathbb{Z}_2$~\cite{Zhou2003,Ueda2014}, is isomorphic to the additive group of integers, $\pi_2(G_{p})\cong\mathbb{Z}$. 

Topological point defects in the polar phase of a $^{87}$Rb spin-1 BEC have been recently realized experimentally~\cite{Ray2015}. No nodal lines or other physical line-like objects are attached to this monopole, and we therefore refer to it as an isolated monopole. As Ref.~\cite{Ray2015} focused only on the first observation of the isolated monopole within a finite experimental resolution, there has been no detailed study of the fine structure of the defect. Furthermore, both the isolated monopole~\cite{Ruostekoski2003} and indeed the entire polar phase of the $^{87}$Rb condensate~\cite{Sadler2006} are expected to be unstable at low magnetic fields, prompting a study on the evolution of the isolated monopole after its creation.


In this Letter, we present computational results of the fine-grained structure and decay dynamics of the isolated monopole. We observe that the core of the created monopole contains a small half-quantum vortex ring as expected based on energetic arguments studied in Ref.~\cite{Ruostekoski2003}. 
We show that beyond the experimentally accessed time scales~\cite{Ray2015} the polar order parameter evolves into a ferromagnetic order parameter, accompanied by the decay of the isolated monopole into a Dirac monopole in the resulting synthetic magnetic field. Importantly, this decay arises naturally from the thoroughly understood physics of the atom cloud without any phenomenological damping terms. Quantitatively matching dynamics are observed for both ferromagnetic and antiferromagnetic spin-spin interactions.

\emph{Theoretical background}---The mean-field order parameter of a spin-1 condensate can be expressed as $\psi(\mathbf{r})=\sqrt{n(\mathbf{r})}\xi(\mathbf{r})$,
where $n(\mathbf{r})$ is the particle density and $\xi(\mathbf{r})$ is a three-component complex-valued spinor such that $\xi(\mathbf{r})^\dagger\xi(\mathbf{r})=1$.
The temporal evolution of the order parameter at low temperatures is accurately described by the differential equation
\begin{align}\label{eq:GP}
i\hbar\frac{\partial }{\partial t}\psi(\mathbf{r}) &= \{ h(\mathbf{r}) + n(\mathbf{r})[c_0 {} \\ \nonumber
{} &+ c_2 \mathbf{S}(\mathbf{r})\cdot\mathbf{F}]- i\Gamma n^{2}(\mathbf{r}) \}\psi(\mathbf{r}),
\end{align}
where $h(\mathbf{r})$ is the single-particle Hamiltonian, $\mathbf{F}=(F_x,F_y,F_z)$ is a vector composed of the dimensionless spin-1 matrices, $\Gamma$ is the three-body recombination rate, and $\mathbf{S}(\mathbf{r})=\xi(\mathbf{r})^\dagger\mathbf{F}\xi(\mathbf{r})$ is the local average spin. The coupling constants characterizing the atom--atom interactions are given by $c_0=4\pi\hbar^2(a_0+2a_2)/(3m)$ and $c_2=4\pi\hbar^2(a_2-a_0)/(3m)$, where $a_f$ is the
$s$-wave scattering length corresponding to the scattering channel with total two-atom hyperfine spin $f$. The single-particle Hamiltonian is given by
\begin{align}
h(\mathbf{r})=&-\hbar^2\nabla^2/(2m)+V(\mathbf{r})-\mu\\ \nonumber
&+g_{\textrm{F}} \mu_{\textrm{B}}\mathbf{B}(\mathbf{r},t)\cdot\mathbf{F}+q[\mathbf{B}(\mathbf{r},t)\cdot\mathbf{F}]^2,
\end{align}
where $m$ is the mass of the atoms, $V(\mathbf{r})$ is an external optical trapping potential, $\mu$ is the chemical potential, $g_F$ is the Land\'{e} $g$-factor, $\mu_B$ is the Bohr magneton, $\mathbf{B}$ is an externally applied magnetic field, and $q$ is the strength of the quadratic Zeeman shift~\cite{Stenger1998}.
We assume that $V(\mathbf{r})=m[\omega_r^2(x^2+y^2)+\omega_z^2 z^2]/2$, where $\omega_r$ and $\omega_z$ are the radial and axial trapping frequencies, respectively.

\begin{figure}[hbtp!]
	\includegraphics[width=0.42\textwidth]{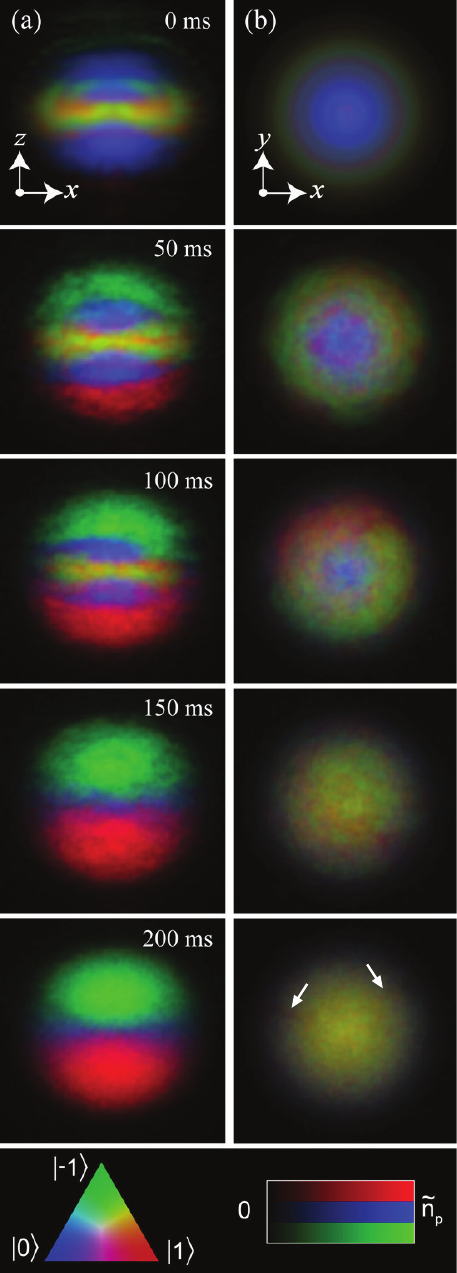}
	\caption{(a) Horizontally (b) and vertically integrated particle densities
for the indicated waiting times after the creation ramp which produces the the isolated monopole. Different colors correspond to particles in different $z$-quantized spin states with the color and intensity
scales given in the bottommost panel. The peak column density is
$\tilde{n}_{\textrm{p}}=2.7\times10^{11}\,\textrm{cm}^{-2}$ and the field of view is $15.5\times 15.5$~$\mu\textrm{m}^2$ in each panel. The white arrows indicate the location of a vortex line shown more clearly in Fig.~\ref{fig:4}}
\label{fig:1}
\end{figure}

In the pure polar phase with $\mathbf{S}(\mathbf{r})=\textbf{0}$, the order parameter can be expressed in the basis of the $z$-quantized spin states $\{\ket{1},\ket{0},\ket{-1}\}$ as~\cite{Mueller2004,Ohmi1998}
\begin{align}\label{eq:order}
\psi(\mathbf{r})=
\frac{\sqrt{n(\mathbf{r})}e^{i\phi(\mathbf{r})}}{\sqrt{2}}
\begin{pmatrix}
-d_x(\mathbf{r})+id_y(\mathbf{r})\\
\sqrt{2}d_z(\mathbf{r})\\
d_x(\mathbf{r})+id_y(\mathbf{r})
\end{pmatrix}_z.
\end{align}
Thus in the Cartesian basis the polar order parameter reads $\mathbf{\psi}(\mathbf{r})=\sqrt{n(\mathbf{r})}e^{i\phi(\mathbf{r})}\hat{\mathbf{d}}(\mathbf{r})$, where $\hat{\mathbf{d}}=(d_x,d_y,d_z)^{{T}}$ is a real unit vector known as the nematic vector. In the case of non-zero average spin, we investigate the nematic order through the magnetic quadrupole tensor \cite{Mueller2004}
\begin{align}\label{eq:Qtensor}
Q_{ab}=\frac{\xi_{a} {\xi_b}^{*} + \xi_{b} {\xi_a}^{*} }{2},
\end{align}
where $\left\{\xi_i\right\}$ are the components of the spinor in the Cartesian basis. For $\langle\mathbf{S}(\mathbf{r})\rangle\neq \textbf{0}$, the vector $\hat{\mathbf{d}}$ is defined as the eigenvector corresponding to the largest eigenvalue of $Q$.


\emph{Methods}---The nematic vector behaves identically to the average spin under rotations in spin space, and hence it also follows adiabatic changes in the external magnetic field. Consequently, the method originally developed in Ref.~\cite{Pietila2009} for the adiabatic creation of Dirac monopoles in the ferromagnetic phase can be used to create isolated monopoles in the polar phase, as realized in Ref.~\cite{Ray2015}. In brief, the condensate is subjected to an external magnetic field $\mathbf{B}(\mathbf{r},t)=\mathbf{B}_{\textrm{q}}(\mathbf{r})+\mathbf{B}_{\textrm{b}}(t)$, where $\mathbf{B}_{\textrm{q}}(\mathbf{r})=b_{\textrm{q}}(x\mathbf{\hat{x}}+y\mathbf{\hat{y}}-2z\mathbf{\hat{z}})$ is a quadrupolar magnetic field with gradient $b_{\textrm{q}}$ and $\mathbf{B}_{\textrm{b}}(t)=B_z(t)\mathbf{\hat{z}}$ is a spatially homogeneous bias field. In the beginning of the simulation, the condensate is in the spin state $\ket{0}$, yielding a nematic vector $\hat{\mathbf{d}}(\mathbf{r})=\hat{\mathbf{z}}$. At the initial bias field $B_z=1$~G, the field gradient is linearly ramped from zero to $b_{\textrm{q}}=3.7$ G/cm in 10 ms and the bias field is subsequently decreased to $B_z=10$ mG in 10 ms. The monopole is created by decreasing the value of the bias field linearly to zero at a rate $\dot{B}_z(t)=-0.25$ G/s. This part of the control protocol is referred to as the \textit{creation ramp} and it ideally results in $\hat{\mathbf{d}}=\hat{\mathbf{B}}_{\textrm{q}}(\mathbf{r})$.
After the creation ramp the temporal evolution continues with the quadrupole field and optical trap intact.

\begin{figure}[hbtp!]
	\includegraphics[width=0.42\textwidth]{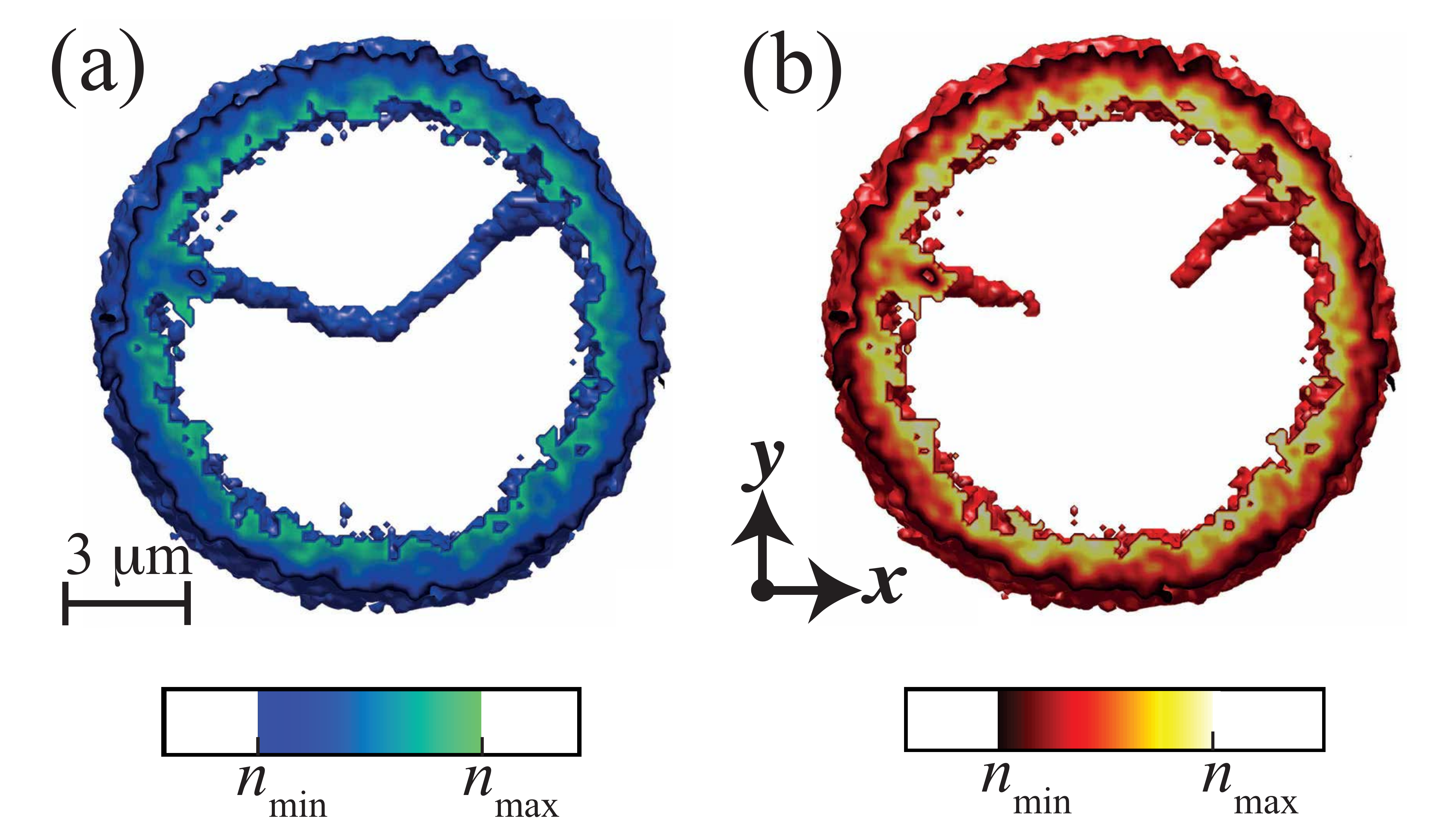}
	\caption{ (a) Spin density $|\psi^\dagger \mathbf{F}\psi|$ and (b) particle density $\psi^\dagger\psi$ of the condensate 200~ms after the creation ramp. The shown density range is $[n_{\textrm{min}}, n_{\textrm{max}}] = [1.0, 4.8]\times 10^{-4}$~$N/a_r^3$.
The data in both panels is shown only for $|z|<1.5a_r$. The spin density is well depleted along the vortex, whereas the particle density is only partially depleted. }
\label{fig:4}
\end{figure}

The initial particle number is $2.1\times 10^5$ and the optical trapping frequencies are $\omega_r =2\pi\times 124$~Hz and $\omega_z=2\pi\times 164$~Hz. 
We take the other parameters according to $^{87}$Rb such as the literature values for the atom loss arising from the three-body recombination $\Gamma=\hbar\times 2.9\times 10^{-30}$~cm$^6$/s~\cite{Burt1997,Stamper-Kurn1998}, the quadratic Zeeman shift $q=2\pi\hbar\times 2.78$~MHz/T~\cite{Sadler2006}, and the scattering lengths $a_0=5.387$~nm and $a_2=5.313$~nm. 
 The computational volume considered is 24$\times$24$\times$24$a_{r}^3$, where $a_r = \sqrt{\hbar /(m\omega_{r})} = 1~\mu$m and the corresponding size of the computational grid is 200$\times$200$\times$200 points.
In order to enhance the numerical emulation of the experimental conditions~\cite{Ray2015}, we add spatially uncorrelated complex noise to the spinor components at each grid point prior to the creation ramp. The amplitude of the noise is uniformly distributed to introduce $0$--$1\%$ fluctuations in the local particle density for each spinor component.

\begin{figure}[hbtp!]
\includegraphics[width=0.42\textwidth]{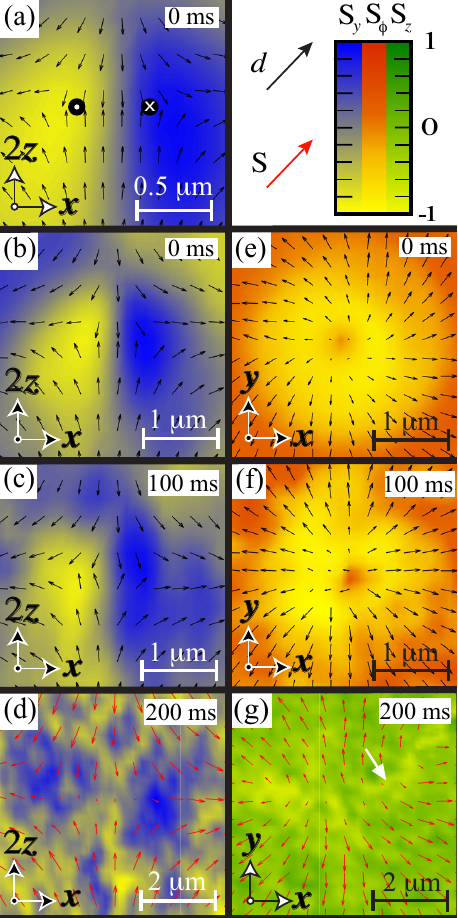}
\caption{(color online) Projection of (a--c, e, f) the nematic vector $\hat{\bf{d}}$ ($\uparrow$) and (d, g) average spin $\hat{\bf{S}}$ (\textcolor{red}{$\uparrow$}) of the condensate  onto (a--d) the $xz$- and (e--g) $xy$-planes for different indicated waiting times after the creation ramp.
The background colormap shows (a--d) the $y$ component, (e,f) the azimuthal
component, and (g) the $z$ component of the spin. The core of the Alice ring is shown with black dots in panel (a) which shows a magnification of the center region of panel (b). The white arrow indicates the location of a vortex line shown more clearly in Fig.~\ref{fig:4} }
\label{fig:2}
\end{figure}

\emph{Results}---We numerically integrate Eq.~\eqref{eq:GP} and apply the control protocol described above with the initial condition $\hat{\mathbf{d}}=\hat{\mathbf{z}}$. Figure~\ref{fig:1} shows the resulting spin-contrast images of the condensate particle densities for different waiting times after the creation ramp. As in Ref.~\cite{Ray2015}, the condensate particle densities just after the creation ramp are in good agreement with Eq.~\eqref{eq:order} and $\hat{\mathbf{d}}=\hat{\mathbf{B}}_{\textrm{q}}(\mathbf{r})$. Thus the particles almost entirely reside in the so-called neutral state which corresponds to the zero eigenvalue of the local Zeeman Hamiltonian.
However, 50~ms after the creation ramp, the polar phase has noticeably decayed towards the local ferromagnetic strong-field seeking state (SFSS), i.e., the spin state that minimizes the local Zeeman energy. The ferromagnetic phase is first visible at the top and bottom edges of the condensate and extends gradually until the condensate resides almost entirely in the SFSS. Qualitatively similar results are obtained in simulations without the added noise (data not shown).


It is well known that the SFSS in the quadrupole field $\mathbf{B}_{\textrm{q}}(\mathbf{r})$ corresponds to the synthetic magnetic field of a Dirac monopole~\cite{Pietila2009}. However, the final order parameter in our case does not have the terminating double-quantum vortex observed in Ref.~\cite{Ray2014}. Instead, there is one single-quantum vortex that reverses its circulation at the monopole, a scenario that has previously been shown to minimize the mean-field energy in the case of a Dirac monopole~\cite{Ruokokoski2011}. We confirm the presence of this single-quantum vortex in Fig.~\ref{fig:4} where it is visible as a line of suppressed spin density. We have verified that the phase winding along this vortex line reverses its sign near the origin where the magnetic field vanishes (data not shown). The orientation of the vortex depends on the particular realization of the applied noise.


Figure~\ref{fig:2} shows the nematic vector and selected components of the spin vector during the decay of the isolated monopole. A ferromagnetic ring with a well-defined polarization is clearly visible just after the creation ramp [see Fig.~\ref{fig:2}(c) and~\ref{fig:2}(d)] although it is so small that it was not distinguished within the finite experimental resolution of Ref.~\cite{Ray2015}. This ring resides at the monopole core and retains its size during the temporal evolution. Since the nematic vector is observed to rotate by $\pi$ about the ferromagnetic core [see Fig.~\ref{fig:2}(a)], the ring is identified as a half-quantum vortex ring, or Alice ring, discussed in Ref.~\cite{Ruostekoski2003}. We determine the radius of the Alice ring from the behavior of the nematic vector to be roughly $0.2$~$\mu$m which exceeds neither the spin healing length $\hbar/\sqrt{2m|c_2|n(0)}\approx 4$~$\mu$m nor the density healing length $\hbar/\sqrt{2m|c_2+c_0|n(0)}\approx 0.3$~$\mu$m. Thus the structure essentially manifests itself a point defect. The subsequent decay of the polar phase destroys the Alice ring and eventually the characterization of the condensate using the nematic vector becomes obscure. We therefore do not show the nematic vector but rather the local spin for long evolution times. Ultimately, the local spin aligns with the external magnetic field as shown in Fig.~\ref{fig:2}(g) and~\ref{fig:2}(h).

\begin{figure}[hbtp!]
	\includegraphics[width=0.45\textwidth]{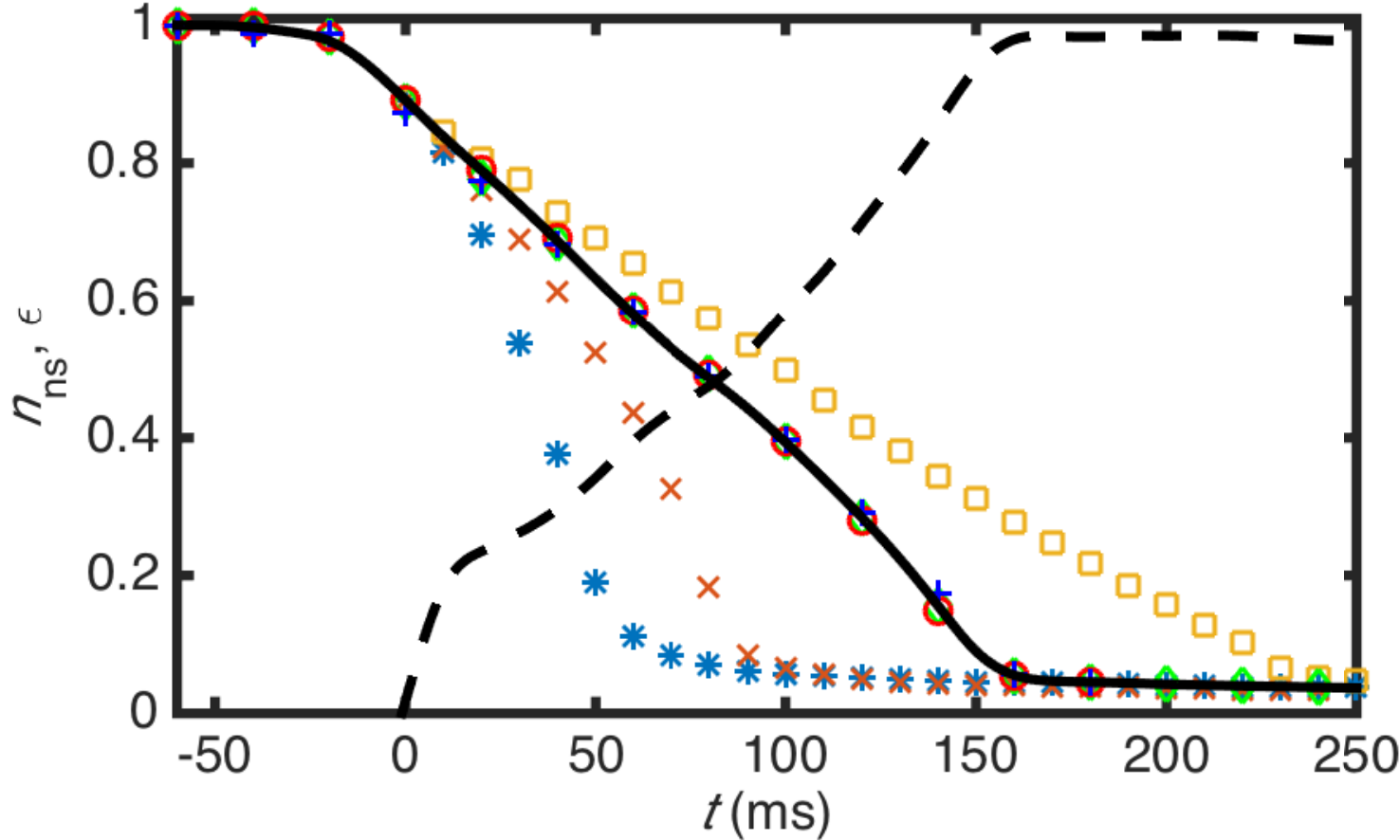}
	\caption{Temporal evolution of the relative population of the neutral state, $n_\textrm{ns}(t)$ (solid line), and deviation of the order parameter from the created isolated monopole at $t=0$, $\epsilon(t)$ (dashed line). In addition to the standard parameters given in the text, neutral state populations are also shown for (i) $q=0$ (\textcolor{red}{$\ocircle$}), (ii) $\Gamma=0$ (\textcolor{blue}{$\textbf{+}$}) and (iii) $c_{2}=-4\pi\hbar^2(a_2-a_0)/(3m)$ (\textcolor{green}{$\diamondsuit$}) in Eq.~\eqref{eq:GP}. Furthermore, we show $n_\textrm{ns}(t)$ for cases, in which the quadrupole field gradient is suddenly changed at $t=0$ from its standard value $b_\textrm{q}=3.7$~G/cm to $1.4\times b_\textrm{q}$ (\textcolor{yellow}{$\Box$}), $0.5\times b_\textrm{q}$ (\textcolor{red}{$\times$}), and $0.3\times b_\textrm{q}$ (\textcolor{blue}{$\ast$}). All curves are interpolated using cubic splines for enhanced visual appearance.}
\label{fig:3}
\end{figure}

Figure~\ref{fig:3} shows the fraction of particles in the neutral state and the deviation of the order parameter from the initial isolated monopole state during the decay. The relative population of the neutral state is given by
$n_\textrm{ns}(t)=\frac{1}{N(t)}\int d\mathbf{r} |\Pi_\textrm{ns}(\mathbf{r},t)\psi(\mathbf{r},t)|^{2}$, where $N(t)$ is the total number of atoms and $\Pi_\textrm{ns}(\mathbf{r},t)$ is a projection to the neutral state~\cite{Ogawa2002}.
The deviation of the order parameter from the initial isolated monopole state at $t=0$ is characterized by $\epsilon(t)=1-|\int d\mathbf{r}\,{\psi}^*(\mathbf{r},t){\psi}(\mathbf{r},0)|/\sqrt{N(0)N(t)}$. We observe that just after the creation ramp, roughly 90\% of the atoms reside in the neutral state, in agreement with the experimentally obtained value in Ref.~\cite{Ray2015}. The decay of the isolated monopole into the ferromagnetic phase is observed to change from an exponential-like behavior into approximately linear decay with increasing magnetic field gradient. This observation suggests that a cascade of decay channels plays a significant role at strong field gradients. Due to the decreasing spatial overlap between the resulting domains with increasing field gradient, the decay dynamics is slower the stronger the field gradient is. For a sufficiently strong gradient, the resulting domains are spatially well separated (see Supplemental Material). 


{Figure~\ref{fig:3} also shows the results obtained for the three additional cases: (i) eliminating the quadratic Zeeman potential, (ii) eliminating three-body recombination, and (iii) reversing the sign of the spin--spin interaction strength. None of these changes leads to a significant effect on the decay dynamics, indicating that the decay is not originating from these terms.
We also studied the creation and decay of the isolated monopole with parameters corresponding to $^{23}$Na atoms and obtained qualitative agreement with the case of $^{87}$Rb atoms (results not shown).


\emph{Conclusions}---Our numerical studies suggest that the isolated monopole structure observed in Ref.~\cite{Ray2015} contains a small Alice ring~\cite{Ruostekoski2003}. This vortex ring is destroyed by a subsequent dynamical phase transition into a ferromagnetic order parameter supporting a Dirac monopole. Although the quadrupole field has been observed to stabilize the polar phase of a $^{87}$Rb condensate if the field zero is well outside the condensate~\cite{Ray2015}, our simulations reveal that after the field zero is brought into the condensate, the polar phase decomposes on a time scale of 100~ms. We attribute this behavior to the spatially varying magnetic field and the linear Zeeman interaction. Neither the spin--spin interactions, quadratic Zeeman effect, nor three-body recombination have a significant effect on the decay dynamics. However, the strength of the magnetic field gradient is shown to have a detrimental effect on the decay speed and characteristics. These studies set the stage for the detailed dynamics of topological point defects in quantum fields. Finding ways to extend the lifetime of the defect further and thereafter to study the dynamics of multiple interacting point defects remain future challenges.


\begin{acknowledgments}
This research has been supported by the Academy of Finland through its
Centres of Excellence Program (Project No.~251748) and Grants No.~135794 and No.~272806, the U.S. National Science Foundation (NSF PHY-1208522),
Finnish Doctoral Programme in Computational Sciences, the Magnus Ehrnrooth Foundation, and the Emil Aaltonen Foundation.
CSC - IT Center for Science Ltd. (Project No. ay2090) and Aalto Science-IT project are acknowledged for computational resources.
\end{acknowledgments}

\bibliographystyle{aipnum4-1}

\bibliography{monopole_decay}

\newpage
\begin{figure}[htp] \centering{
\includegraphics[scale=0.9]{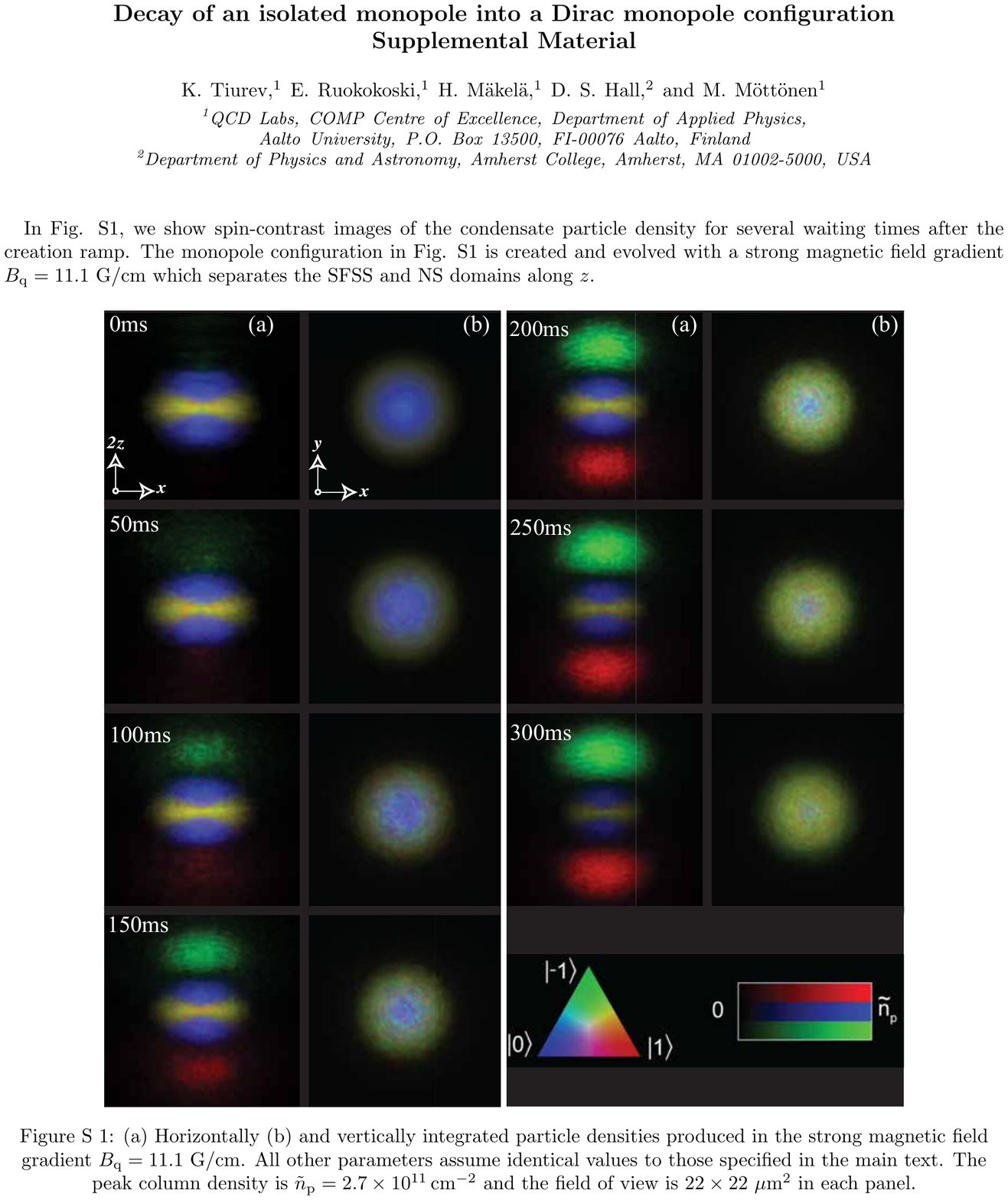}}
\end{figure}

\end{document}